\documentclass[aip,jap,reprint,floatfix]{revtex4-1}

\pdfoutput=1
\usepackage{amsfonts, amssymb, amsbsy, amsmath}
\usepackage{graphicx}
\usepackage{epstopdf}
\usepackage{array,multirow}

\newcommand{\figta}{$\left(\mathrm{a}\right)\;$}
\newcommand{\figtb}{$\left(\mathrm{b}\right)\;$}
\newcommand{\figtc}{$\left(\mathrm{c}\right)\;$}

\newcommand{\figa}{$\left(\mathrm{a}\right)$}
\newcommand{\figb}{$\left(\mathrm{b}\right)$}

\newcommand{\mk}{\;\mr{mK}}
\newcommand{\nm}{\;\mr{nm}}
\newcommand{\kelvin}{\;\mr{K}}
\newcommand{\na}{\;\mr{nA}}
\newcommand{\mev}{\;\mr{meV}}
\newcommand{\volt}{\;\mr{V}}
\newcommand{\ohm}{\;\Omega}
\newcommand{\kohm}{\;\mr{k}\Omega}
\newcommand{\muv}{\;\mu\mr{V}}
\newcommand{\mua}{\;\mu\mr{A}}

\newcommand{\mr}[1]{\mathrm{#1}}
\newcommand{\kb}{k_{\mr{B}}}
\newcommand{\tc}{T_{\mr{c}}}
\newcommand{\lmin}{L_1}
\newcommand{\lmax}{L_2}

\newcommand{\rsquare}{R_{\square}}
\newcommand{\reff}{R_{\mr{eff}}}
\newcommand{\rn}{R_{\mr{N}}}
\newcommand{\rt}{R_{\mr{T}}}

\newcommand{\isw}{I_{\mr{sw}}}
\newcommand{\iswmax}{I_{\mr{sw}}^{\mr{max}}}
\newcommand{\ir}{I_{\mr{r}}}
\newcommand{\isns}{I_{\mr{SNS}}}
\newcommand{\vsns}{V_{\mr{SNS}}}
\renewcommand{\eth}{E_{\mr{Th}}}
\newcommand{\didphimax}{|\partial{I}/\partial\Phi|_{\mr{max}}}
\newcommand{\didphi}{\partial{I}/\partial\Phi}
\newcommand{\iphi}{I(\Phi)}

\begin{document}

\title{Low-temperature characterization of Nb-Cu-Nb weak links with Ar ion-cleaned interfaces}

\author{R. N. Jabdaraghi}
\email{robab.najafi.jabdaraghi@aalto.fi}
\affiliation{Low Temperature Laboratory, Department of Applied Physics, Aalto University School of Science,
P.O. Box 13500, FI-00076 Aalto, Finland}
\affiliation{Faculty of Physics, University of Tabriz, Tabriz, Iran}

\author{J. T. Peltonen}
\affiliation{Low Temperature Laboratory, Department of Applied Physics, Aalto University School of Science,
P.O. Box 13500, FI-00076 Aalto, Finland}

\author{O.-P. Saira}
\affiliation{Low Temperature Laboratory, Department of Applied Physics, Aalto University School of Science,
P.O. Box 13500, FI-00076 Aalto, Finland}

\author{J. P. Pekola}
\affiliation{Low Temperature Laboratory, Department of Applied Physics, Aalto University School of Science,
P.O. Box 13500, FI-00076 Aalto, Finland}

\begin{abstract}
We characterize niobium-based lateral Superconductor (S) -- Normal metal (N) -- Superconductor weak links through low-temperature switching current measurements and tunnel spectroscopy. We fabricate the SNS devices in two separate lithography and deposition steps, combined with strong argon ion cleaning before the normal metal deposition in the last step. Our SNS weak link consists of high-quality sputtered Nb electrodes that are contacted with evaporated Cu. The two-step fabrication flow enables great flexibility in the choice of materials and pattern design. A comparison of the temperature-dependent equilibrium critical supercurrent with theoretical predictions indicates that the quality of the Nb-Cu interface is similar to that of evaporated Al-Cu weak links. Aiming at increased sensitivity, range of operation temperatures, and thermal isolation, we investigate how these SNS structures can be combined with shadow-evaporated aluminum tunnel junctions for sensor applications that utilize the superconducting proximity effect. To this end, we demonstrate a hybrid magnetic flux sensor based on a Nb-Cu-Nb SNS junction, where the phase-dependent normal metal density of states is probed with an Al tunnel junction.
\end{abstract}

\date{\today}

\maketitle

A Superconductor -- Normal metal -- Superconductor (SNS) weak link~\cite{Tinkham,Likharev} consists of a short normal conducting metal (N) embedded between two superconducting (S) electrodes and supports supercurrent due to proximity effect.~\cite{Tinkham,Gennes66,Gennes64,Buzdin05,Belzig99,Pannetier00}
Lateral SNS weak links allow the creation of further novel types of interferometers~\cite{Petrashov95,Giazotto10} with a large number of foreseen applications including measurement of magnetic flux induced by atomic spins, single-photon detection and nanoelectronical measurements.~\cite{Foley,Hao05,Hao07} In many instances, the performance of these sensors improves when the proximity effect in the N wire, as measured by the magnitude of the induced (mini)gap, is maximized. In this respect, moving to a large-gap superconductor such as Nb or Nb(Ti)N is an evident direction to look for performance gains.

\begin{figure}[htb]
\includegraphics[width=\columnwidth]{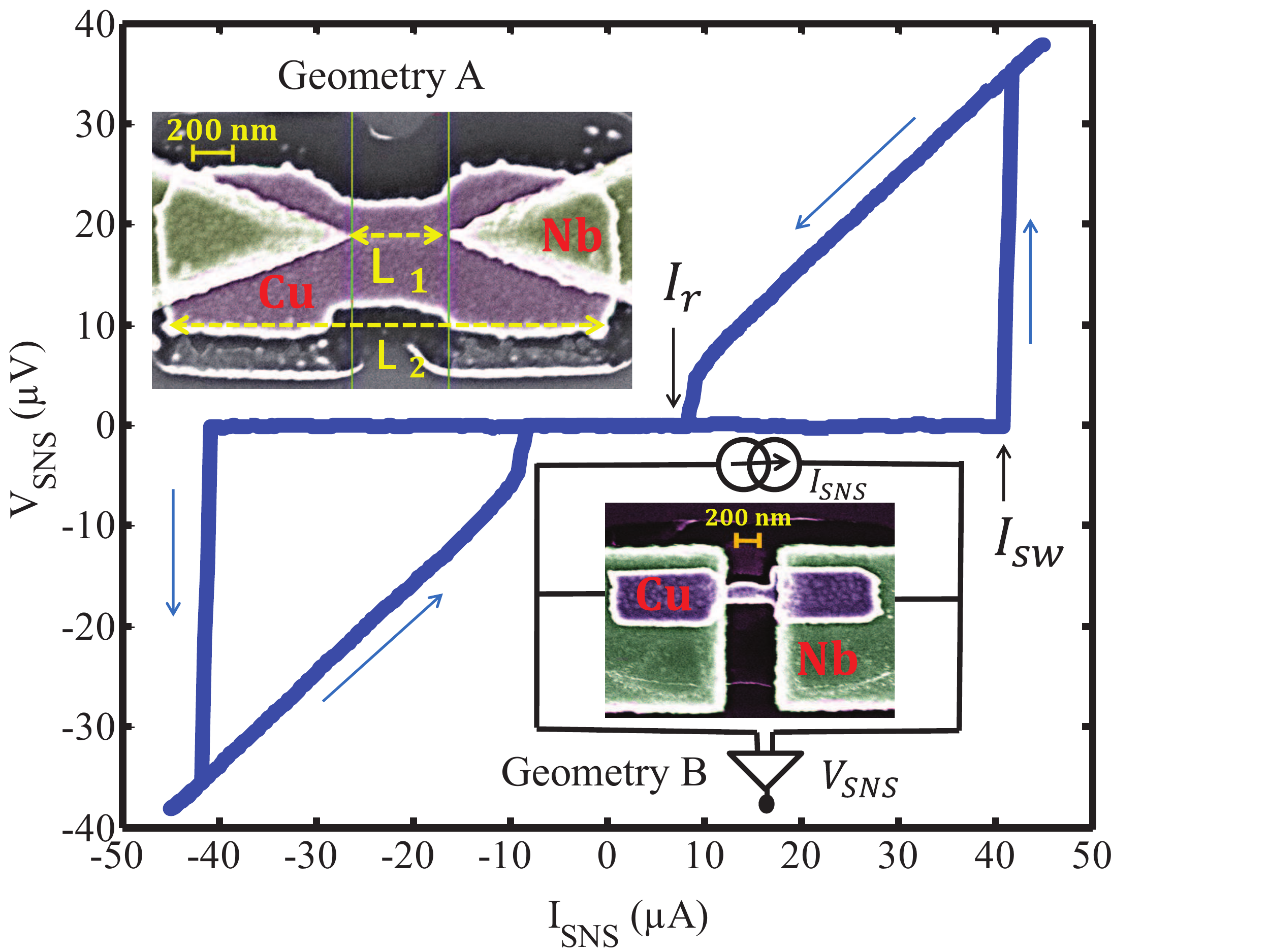}
\caption{Main panel: current--voltage characteristic of sample B2 measured at $T=80\mk$. Arrows indicate the switching $\isw$  and retrapping current $\ir$. The top (bottom) inset shows a representative scanning electron micrograph of a junction with geometry A (B), consisting of a normal metal (Cu) island embedded between two superconducting electrodes (Nb). The distance between the superconducting electrodes is $\lmin$, whereas $\lmax$ is the total N  length. The bottom inset includes a sketch of the measurement setup.}
\label{fig:sample}
\end{figure}

Aluminium-based weak links with well controlled interface transparencies are routinely fabricated by shadow evaporation through a bilayer mask~\cite{Dolan77} that allows deposition of the superconducting and normal electrodes in a single vacuum cycle. When it comes to superconductors with higher $\tc$ compared to Al, vanadium is one of the few that are easily suited for shadow evaporation.~\cite{pascualgarcia09,ronzani13} On the other hand, in Ref.~\onlinecite{Dubos2001} the critical currents in shadow-evaporated Nb-Cu-Nb SNS junctions, measured down to $300\mk$, were found to be in excellent agreement with theoretical predictions. However, the evaporation of good quality Nb films requires significant attention due to the high melting temperature and the resulting organic resist outgassing: Either a bilayer mask with a special thermostable polymer~\cite{Dubos2000,Chiodi09}, a fully inorganic mask~\cite{hoss99,samaddar13}, or an evaporator with large target-to-sample distance~\cite{Nam02} has to be used.

For many detector applications of SNS weak links such as sensitive hot-electron bolometers and calorimeters~\cite{govenius14,gasparinetti15}, it can be moreover desirable to avoid any unwanted normal metal structures, including the shadow replicas always present after multi-angle evaporation. An etching-based, two-step process would also be preferable on the grounds of unrestricted geometry when combining proximity weak links with high-quality superconducting resonators patterned from Nb, Ta, or other difficult-to-evaporate materials. In addition, if such Nb structures are to be combined with shadow evaporated structures and tunnel junctions, it is advantageous to start with the Nb deposition: The tunnel junctions or merely individual thin films of several materials are likely to show degradation if they are not formed in the last process step. An alternative approach would be to start by fabricating the normal metal electrode from a noble metal, e.g., gold, whose surface is relatively easy to clean, and to deposit the Nb electrode only in the second step. However, due to added complexity, this may not be optimal in combining such an SNS weak link with shadow evaporated tunnel probes in further steps. 

In this Letter, we report a study of sputtered niobium -based SNS weak links where the normal metal electrode is deposited shortly after an \emph{in situ} argon ion etching. Similar junctions with Ar-cleaned interfaces have been fabricated before~\cite{Dubos2001,baselmans01,baselmans02,baselmans02b}, but a detailed investigation in a simple SNS geometry has not been reported. We consider the structures from the particular viewpoint of combining them with shadow-evaporated normal and superconducting metals. The S--N interface quality can be verified by comparing the temperature dependence of the measured critical (switching) current with a theoretical model valid in the diffusive limit. To study how to achieve minimum S electrode separation with good fabrication yield, we consider both triangular and rectangular shaped terminations (see Fig.~\ref{fig:sample}, insets), referred to as geometry A and B, respectively. The thickness profile at the tip of the electrode is more gentle for geometry A, which may affect the contact quality and the extent of inverse proximity effect. As an application of the transparent contacts, we report the first experimental demonstration of a Nb-based SQUIPT (superconducting quantum interference proximity transistor) interferometer.~\cite{Giazotto10} In an initial device we observe maximum flux-to-current transfer function values of about $\didphimax=50\na/\Phi_0$ at $T=80\mk$. Here, $\Phi_0=h/(2e)$ is the superconducting flux quantum.

\begin{table}[htb]
\caption{Parameters of measured samples. $\lmin$ is the minimum Nb electrode separation in the SNS junction, $\lmax$ the full length of Cu wire, and $w$ the minimum width of the copper island (see Fig.~\ref{fig:sample}). $\eth$, $\reff$ and $L$ are obtained from a comparison of the temperature-dependent switching current measurements to the theoretical model. $\rn$ is the measured low-temperature normal state resistance of the wire, and $\iswmax$ denotes the maximum observed switching current.}
\centering

\begin{tabular}{ccccccccccc}
\hline\hline

\parbox[t]{2mm}{\multirow{2}{*}{\rotatebox[origin=c]{90}{{\tiny sample}}}} & $\lmin$ & $\lmax$ & $w$ & $L$ & $\reff$ & $\rn$ & $\eth$ & $\iswmax$ & \multirow{2}{*}{$\frac{e\reff\iswmax}{\eth}$}  \\

& ($\mu\text{m})$ & ($\mu\text{m})$ & ($\mu\text{m})$ & ($\mu\text{m})$ & $(\Omega)$ &  $(\Omega)$ &$(\mu\text{eV})$ & ($\mu\text{A})$ & & \\

\hline
A1 & 0.29 &2& 0.55 &1.09 &1.6& 0.83& 5.5& 33&9.6\\
A2 & 0.36 &2 &0.55 &1.18 &1.25& 0.83&4.7&33&8.8\\
A3 & 0.48 &2.15&0.54 & 1.18 &1.2&0.57& 4.8&30&7.5\\
A4 & 0.49 &2.15&0.54 &1.21 &1.2& 0.59&4.5&30&8\\
A5 & 0.84 &2.5&0.85 &1.80 &1.03&0.5&2& 10&5.2\\
B1 & 0.32 &2&0.15 &1.09 &1.6& 0.98& 7.7& 42&8.7\\
B2 & 0.38 &2&0.15 &1.09 &1.55&0.89&7.3& 41&8.7\\
B3 & 0.40 &2.12&0.15 &0.96 &4.6& 2.31&7& 14&9.2\\
B4 & 0.39 &2.17&0.14 &0.91 &5.2&2.42&8& 14&9.1\\
B5 & 0.40 &2.12&0.15 &0.85 &4.7&2.51&9& 18&9.4\\
\hline\hline
\end{tabular}
\label{tab:samples}
\end{table}

The fabrication process consists of two main steps: First making the Nb structures and then providing clean electric contact between Nb and Cu. The starting point is an oxidized 4 inch Si substrate with $200\nm$ sputter-deposited Nb. An etch mask based on positive tone AR-P 6200.13 resist is prepared by electron-beam lithography (EBL), followed by wet development and 5 min reflow baking at $150^{\circ}\mr{C}$ to avoid abrupt edge profiles in the Nb structures. Reactive ion etching (RIE) with a mixture of $\mr{SF}_6$ and Ar is then used to transfer the pattern into the Nb film. We employed gas flows of 20 sccm and 10 sccm for $\mr{SF}_6$ and Ar, respectively, and the 100 W RF power resulted typically in DC self-bias of $300-320\volt$. Nb etching is followed by a second round of EBL using a conventional bilayer resist.~\cite{Dolan77} In our case this consists of a $200\nm$ ($600\nm$) thick poly(methyl methacrylate) layer for geometry A (B), on top of a $900\nm$ layer of copolymer.

The crucial step in the fabrication is to create a transparent contact between Nb and Cu. This is done by exposing the chip to \emph{in situ} Ar ion etching, in the same vacuum cycle immediately prior to the Cu deposition. Based on profilometer traces we estimate that typically $10-20\nm$ of Nb is removed in this cleaning step. To complete the SNS junction, the $60\nm$ thick normal metal Cu electrode is deposited by electron-gun evaporation. We find typical low-temperature values of the Cu sheet resistance $\rsquare$ to be in the range $0.4-0.5\ohm$. The Ar ion flux is meeting the sample perpendicular to the substrate, same as the deposited Cu.

\begin{figure}[htb]
\includegraphics[width=\columnwidth]{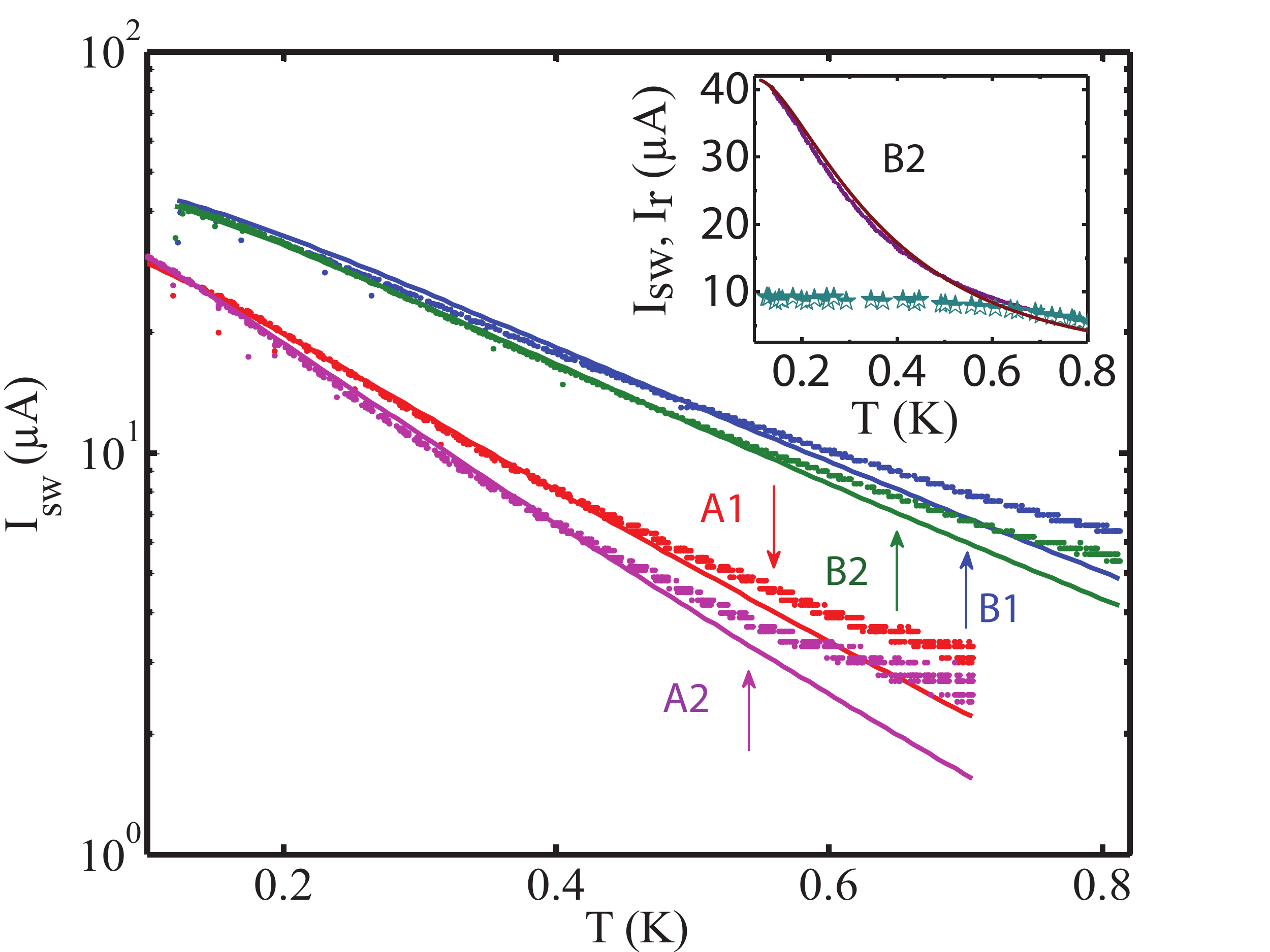}
\caption{The symbols show the temperature dependence of the measured $\isw$ of samples A1, A2, B1, and B2. The solid lines show predictions of a simple model based on the Usadel equations [Eq.~(\ref{eq:ic})]. The adjustable parameters are the Thouless energy $\eth$ and effective wire resistance $\reff$. The vertical arrows identify the samples and indicate the temperature below which the hysteresis in the IV curve appears for each of them. Inset: Temperature dependence of the measured $\isw$ together with the theoretical model (solid lines) for sample B2. The star symbols show the rerapping current $\ir$, which is almost temperature-independent in the hysteretic regime.} \label{fig:isw}
\end{figure}

The main panel of Fig.~\ref{fig:sample} shows the current--voltage characteristic of one of the measured samples, SNS junction B2, featuring a well-defined supercurrent. All the measurements were performed in a dilution refrigerator, down to the base temperature between $50-100\mk$. As indicated in the bottom inset of Fig.~\ref{fig:sample}, the weak link is biased by a current $\isns$ and the dc voltage $\vsns$ is measured in a four-probe configuration. In this sample we observe hysteretic behavior~\cite{Courtois2008,Chiodi09} at $T<0.65\kelvin$ from which we deduce a switching current $\isw$ and retrapping current $\ir$ of about $42\mua$ and $9\mua$, respectively.

The parameters of the measured junctions are listed in Table~\ref{tab:samples}. The symbols in Fig.~\ref{fig:isw} summarize the measured switching currents $\isw$ as a function of the bath temperature $T$ for four of the samples. For clarity, the inset of Fig.~\ref{fig:isw} further shows the temperature dependence of the switching and retrapping currents for sample B2 on a linear scale. We find no qualitative differences in the behavior of the switching currents between junction geometries A and B. The rectangular geometry B, however, has higher fabrication yield for junctions with short $\lmin$, despite the steeper edge profile.

In diffusive metallic SNS junctions the natural energy scale for the proximity effect is given by the Thouless energy $\eth=\hbar D/L^2$, where $D$ and $L$ are the diffusion coefficient and the length of the normal metal electrode, respectively. Our samples are in the long junction limit $L\gg\xi_0$, or equivalently $\Delta\gg\eth$, where $\xi_0=(\hbar D/\Delta)^{1/2}$ and $\Delta\approx 1.2\mev$ denote the superconducting coherence length and energy gap, respectively. The solid lines in Fig.~\ref{fig:isw} are calculated according to
\begin{equation}
\isw=\frac{\eth}{2e\reff}\int_{-\infty}^{\infty} dE\,\mr{Im}\left[j_{E}(E)\right]\tanh\left(\frac{E}{2\kb T}\right).\label{eq:ic}
\end{equation}
Here, the energy-dependent quantity $j_{E}(E)$, depending only weakly on the ratio $\Delta/\eth$ in the long junction limit, is a spectral supercurrent obtained from a numerical solution of the Usadel equations in a 1D SNS geometry, assuming perfectly transparent interfaces~\cite{Dubos2001,Tero2002}. As adjustable parameters we use the Thouless energy $\eth$ and an effective resistance $\reff$. The values resulting in best agreement are listed in Table~\ref{tab:samples} for all samples. The effective junction lengths $L$, derived from the estimated values of $\eth$ and the measured Cu diffusion constant $D\approx 0.01\;\mr{m}^{2}\mr{s}^{-1}$, satisfy $\lmin<L<\lmax$ as expected. Moreover, the ratios $\rn/\reff\approx 0.5$ are similar to our shadow-evaporated Al-Cu weak links~\cite{Courtois2008}, indicating transparent nature of the interfaces.

The simple model with a single temperature-independent effective length and effective resistance is not able to capture the behavior of the critical current at the highest temperatures, evident as changing slope of the experimental curves on the semilogarithmic scale. This can be partially explained by the transition of the IV curves from hysteretic to nonhysteretic, combined with our definition of the switching current as the bias current showing maximum $|\partial\vsns/\partial\isns|$, employed over the full temperature range. Depending on the sample geometry, we obtain values for the effective lengths $L/\xi_0\approx 15-20$, corresponding to $\Delta/\eth\approx 130-600$. The values of $L$ are larger than $\lmin$ but smaller than $\lmax$ due to inverse proximity effect where the Cu island overlaps the Nb electrodes. In the model we neglected the self-consistency and temperature dependence of the order parameter as well as assumed a sinusoidal current-phase relation.~\cite{Tero2002,Sueur08}

\begin{figure}[htb]
\includegraphics[width=\columnwidth]{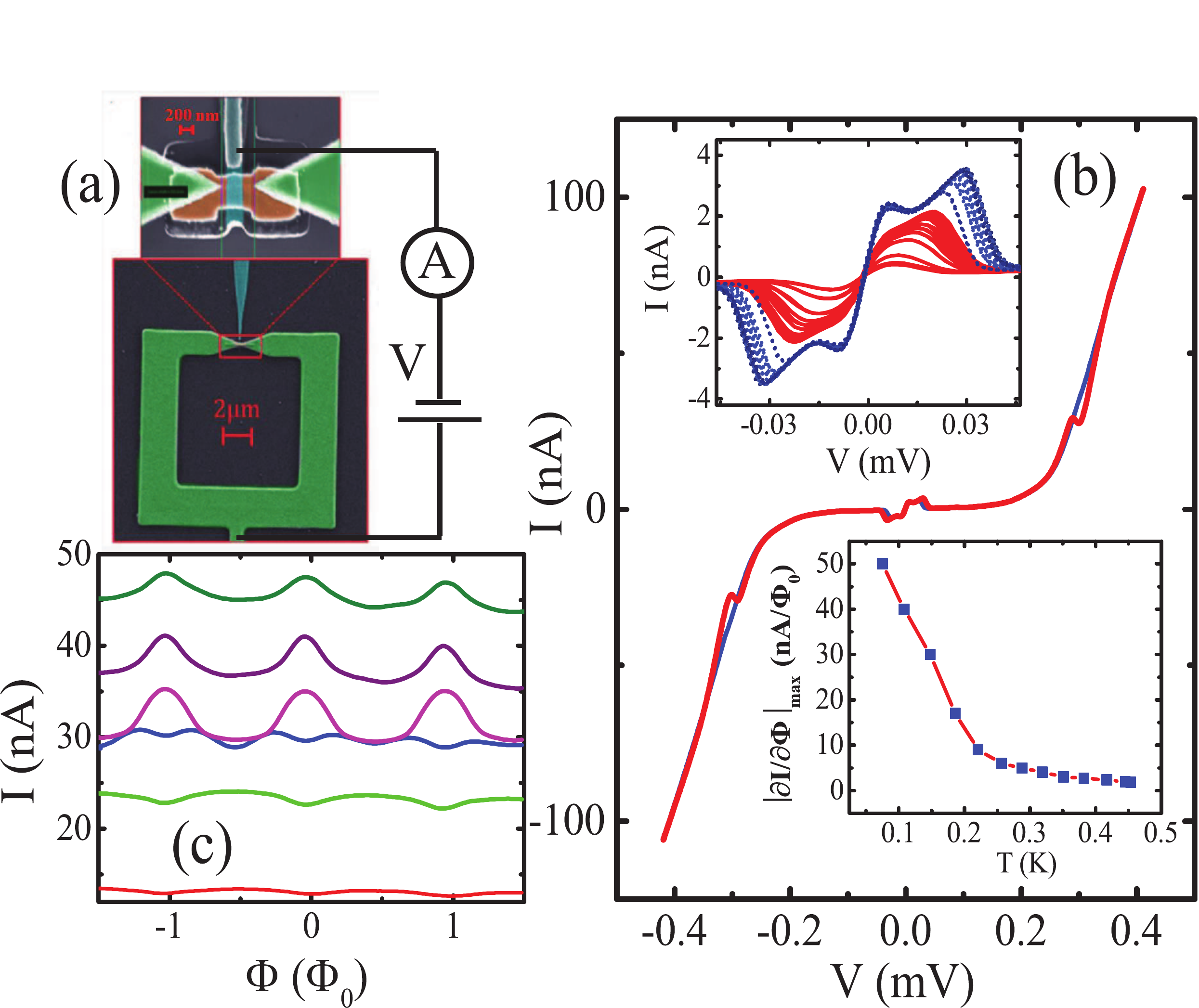}
\caption{\figta False color scanning electron micrograph of an initial Nb-SQUIPT, together with the scheme for measurement under voltage bias. The enlarged view emphasizes the Al superconducting tunnel probe (cyan) connected to the middle of the Cu island (brown) in the Nb-Cu-Nb weak link. Nb electrodes and the loop are shown in green. \figtb IV characteristics of a Nb-SQUIPT based on an A-type SNS junction measured at $T=80\mk$, for two values of magnetic flux through the loop: $\Phi=0$ (red) and $\Phi=0.5\Phi_0$ (blue). The upper inset shows the flux modulation of the IV curve around zero bias voltage at $T=190\mk$ (red solid) and  $T=80\mk$ (blue dashed) while the lower inset presents temperature dependence of the maximum flux-to-current transfer function $\didphimax$ on the supercurrent branch. \figtc Current modulation $\iphi$ at $T=80\mk$ for several values of the bias voltage between 250 (bottom curve, red) and $320\muv$ (top curve, dark green).}
\label{fig:squipt}
\end{figure}

In samples A5 and B3--B5 the normal metal Cu electrode was deposited immediately after the Ar cleaning. With the rest of the samples we investigated combining the Nb-Cu-Nb weak link with an Al-AlOx-Cu Normal metal -- Insulator -- Superconductor (NIS) tunnel junction. To this end, structures A1--A4 and B1--B2 were fabricated simultaneously with SQUIPT-interferometers (see the following paragraphs and Fig.~\ref{fig:squipt}) located elsewhere on the same chip. In these SNS junctions, after Ar cleaning but prior to Cu deposition, the Nb contact areas were controllably exposed to an atmosphere of pure oxygen, typically for 1--5 minutes at a pressure of 1--5 mbar. Importantly, based on values of $\rn$ as well as the ratios $\rn/\reff$, we conclude that this \emph{in situ} oxidation of the exposed Nb contact surfaces has less influence on the SNS weak link properties than variation between different fabrication rounds in the alignment of the Cu island, the Cu deposition, and the nominally identical Ar etching conditions.

Imposing a phase difference between the superconducting electrodes of a SNS junction provides a means to modulate the density of states (DoS) in the N electrode and the supercurrent through the weak link.~\cite{Petrashov94,Petrashov95,Belzig02} A SQUIPT interferometer,~\cite{Giazotto10} see Fig.~\ref{fig:squipt}~\figa, consists of an SNS link embedded into a superconducting loop. When placed into perpendicular magnetic field, the flux-dependent density of states can be accurately probed with a weakly coupled tunnel junction in contact with the N electrode. The performance of such hybrid devices with Al-Cu weak links has improved as a result of several studies.~\cite{Giazotto10,Meschke11,Najafi14,Ronzani14,Giazotto11} To realize an Nb-SQUIPT with our present technique, we first prepare a suitable shadow mask and Ar etch the Nb contacts, in a fashion identical to the Nb-Cu-Nb weak links above. Then, without breaking the vacuum, immediately before the Cu evaporation, a $20\nm$ thick Al electrode of the NIS junction is deposited and subjected to \emph{in situ} oxidation to create the thin AlOx tunnel barrier.

Figure~\ref{fig:squipt}~\figtb displays the IV characteristics of the device measured at two different magnetic fields, $\Phi=0$ and $\Phi=0.5\Phi_0$, corresponding to maximum and minimum minigap opened in the normal metal.~\cite{Sueur08,Zhou98} Due to the relatively low tunnel junction resistance $\rt\approx 3\kohm$, close to zero voltage bias we observe a maximum supercurrent of $3.6\na$ at $T=80\mk$. The overall shape of the characteristics follows that of the quasiparticle tunneling in a single NIS junction, with the addition of a flux-dependent onset of current when $|V|$ exceeds the sum of the S electrode gap $\Delta/e$ and the minigap induced in the N wire. The solid red lines in the upper inset of Fig.~\ref{fig:squipt}~\figtb demonstrate the extent of the flux modulation of the IV curve around zero bias at $T=190\mk$, where we observe almost full closing of the minigap. For comparison, the IVs at $T=80\mk$ are included as the dashed blue lines.

To characterize the flux sensitivity of the SQUIPT device, we measured $\iphi$ curves at several values of the constant bias voltage $V$ and subsequently obtained the flux-to-current transfer function $\didphi$ by numerical differentiation. Figure~\ref{fig:squipt}~\figtc shows some of the flux modulations $\iphi$ measured at $V$ around the onset of the quasiparticle current. In this range of bias voltages, we find the maximum sensitivity $\didphimax\approx40\na/\Phi_0$. Interestingly, for this device the overall maximum $\didphimax\approx50\na/\Phi_0$, at the base temperature around $80\mk$, is reached in the supercurrent branch at $V\approx 32\muv$. For comparison, $\didphimax\approx100\na/\Phi_0$ at $T=240\mk$ has been recently reported for an optimized Al-SQUIPT.~\cite{Ronzani14} The temperature dependence of the maximum sensitivity in the low-bias regime is further shown in the bottom inset of Fig.~\ref{fig:squipt}~\figb, decreasing monotonously as $T$ increases.

The most significant improvements to the performance of this initial realization of a Nb-SQUIPT are expected to result from decreasing the effective length $L$ of the SNS weak link closer to the intermediate-to-short junction regime. We estimate that the present etching and lithography techniques allow $\lmin$ to be reliably reduced down to $100\nm$ in both studied lead geometries, in particular with somewhat thinner Nb electrodes. For consistent values of $\lmin$ the rectangular geometry is preferable. Together with shorter Nb-Cu overlap lengths, i.e., $\lmax$ closer to $\lmin$, this would bring the theoretical performance estimates~\cite{Giazotto10,Giazotto11} for an Nb-SQUIPT with $L=150\nm$ within reach. The sensitivity may further increase with narrower Al probe junctions leading to less spatial averaging of the N DoS, as a result of optimization of the Ar cleaning step and the shadow mask. In the present device we observe a notable broadening of the features after the \emph{in situ} etching.

In conclusion, we have investigated Nb-Cu-Nb weak links based on two independent lithography and deposition steps, relying on Ar ion cleaning of the Nb contact surfaces. The work helps improving the performance of superconducting magnetometers, as well as in including submicron lateral SNS weak links in other detector applications, such as ultrasensitive bolometers and calorimeters, where they need to be integrated with shadow-evaporated tunnel junctions.
\\

The work has been supported by the Academy of Finland Centre of Excellence program (project number 250280). We acknowledge the Iranian Ministry of Science, Research and Technology, and thank Micronova Nanofabrication Centre of Aalto University for providing the processing facilities. We thank L. Gr\"onberg for help with Nb deposition, and D. Golubev, M. Meschke and P. Virtanen for helpful discussions.

\end{document}